\documentclass[usenatbib]{mn2e}
\usepackage{fleqn,graphicx,amssymb}
\usepackage{multirow}
\renewcommand{\[}{\begin{equation}}
\renewcommand{\]}{\end{equation}}
\def\pa{\partial}\def\Henon{H\'enon}
\def\Lc{L_{\rm c}}
\def\figref#1{Fig.~\ref{#1}}
\def\df{{\sc df}}
{\newif\ifnotend
\notendtrue
\def\veclist{ABCDEFGHIJKLMNOPQRSTUVWXYZabcdefghijklmnopqrstuvwxyz.}
\def\top#1#2.{#1}
\def\tail#1#2.{#2.}
\loop\expandafter\xdef\csname v\expandafter\top\veclist\endcsname%
{{\noexpand\bf\expandafter\top\veclist}}
\edef\veclist{\expandafter\tail\veclist}
\if\veclist.\notendfalse\fi\ifnotend\repeat}
\def\d{{\rm d}}

\def\df{\textsc{df}}

\def\d{\mathrm{d}}

\def\fracj#1#2{{\textstyle{#1\over#2}}}

\title[Henon's Isochrone Model]
{Henon's Isochrone Model}

\author[James~Binney]{
  James~Binney
  \thanks{E-mail: j.binney1@physics.ox.ac.uk}\\  
  Rudolf Peierls Centre for Theoretical Physics, 1 Keble Road,
  Oxford, OX1 3NP, UK
}

\begin{document}
\maketitle

\begin{abstract}
\Henon\ sought the most general spherical potential in which the radial
periods of orbits depended only on energy. He named this potential the
isochrone, and discovered that it provided a good representation of data for
globular clusters. He sought an explanation in terms of resonant relaxation.
The role that resonant relaxation might play in globular clusters is still an
open question, but the isochrone potential is guaranteed a role in dynamical
astronomy because it is the most general potential in which closed-form
expressions for angle-action coordinates are available. I explain how this
property makes the isochrone invaluable for the powerful technique of torus
mapping. I also describe flattened isochrone models, which enable us to explore a
powerful general method of generating self-consistent stellar systems.
\end{abstract}

\begin{keywords}
  solar neighbourhood -- Galaxy: kinematics and dynamics -- methods:
  data analysis
\end{keywords}

\section{Introduction} \label{sec:intro}

I have been asked to speak about \Henon's isochrone model. This was covered
in three papers, submitted to Ann Ap in French between May 1958 and November
1959 \citep{HenonI,HenonII,HenonIII}. The first of these papers is by far the
most non-trivial. In it \Henon\ derives the isochrone potential as the most
general spherical potential in which the radial period is independent of
angular momentum at a given energy. Since this paper appeared in French and
has a subtle line of argumentation I have decided to allocate a good deal of
space to reproducing, and I hope clarifying, its argument. The other two
papers are much more straightforward and I will not cover them in detail.
Instead I will describe the impact that the isochrone model has had on my
research over many years. In particular I will describe the role the
isochrone potential plays in torus mapping, and a family of flattened
isochrone models which I have recently published. I believe these
applications will ensure that \Henon's isochrone plays a significant role in
astrophysics throughout the coming decade.

\section{The first isochrone paper}

The paper of May 1958 \citep{HenonI} points out that the gravitational
potentials of both a homogeneous mass distribution and a point mass have the
property that the radial period $T_r$ of an orbit depends on the orbit's
energy $E$ alone: all orbits of a given value of $E$ have the same value of
$T_r$, irrespective of their angular momentum $L$. \Henon\ asks ``what is the
most general gravitational potential for which this property holds?''

\begin{figure}
\centerline{\includegraphics[width=.8\hsize]{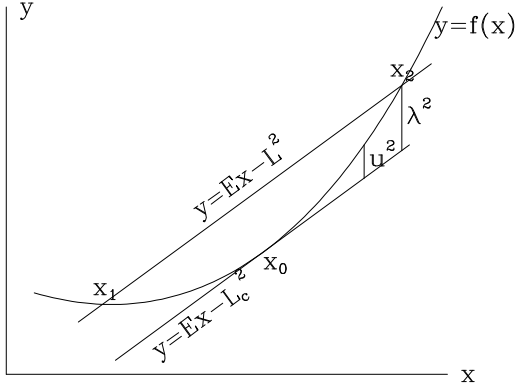}}
\caption{Curves in the plane of $x=2r^2$ and $f=x\Phi(r)$.}\label{fig:one}
\end{figure}

He writes down the integral $\int \d r/v_r$ that determines $T_r$ and makes a
change of variable to $x=2r^2$. Then the integral becomes 
\[\label{eq:Tr}
T_r=\int_{x_1}^{x_2}{\d x\over\sqrt{Ex-L^2-f(x)}},
\]
 where $f(x)\equiv x\Phi[r(x)]$. He plots the curve $y=f(x)$ defined by the
potential and points out that the argument of the radical on the bottom of
(\ref{eq:Tr}) is the vertical distance between this curve and the straight
line $y=Ex-L^2$.  Increasing $L$ has the effect of moving the straight line
down, and at a critical value $\Lc(E)$ the line touches the
convex curve $y=f(x)$ at $x_0$ (\figref{fig:one}). \Henon\
introduces a new independent variable $u$ through
 \[
u^2\equiv f(x)-(Ex-\Lc^2)
\]
 along with the convention that $u<0$ when $x<x_0$ and $u>0$ when $x>x_0$.
Geometrically, $u^2$ is the vertical distance between the tangent $y=Ex-\Lc^2$ and the
the curve $y=f(x)$. For $L<\Lc$ the line $y=Ex-L^2$ intersects the curve
$y=f(x)$ twice: at $x_1<x_0$ and at $x_2>x_0$. These points of intersection
are implicitly defined by the equation
 \[\label{eq:d1}
Ex_i-L^2-f(x_i)=0\quad(i=1,2).
\]

The quantity
 \[\label{eq:defslam}
\lambda^2=\Lc^2-L^2
\]
 is the vertical distance between the two straight lines in \figref{fig:one},
and at $x_1$ $u=-\lambda$ and at $x_2$ $u=\lambda$.

With these definitions the integral for $T_r$ can be rewritten
 \[
T_r=\int_{x_1}^{x_2}{\d x\over\sqrt{Ex-\Lc^2+\lambda^2-f(x)}}
=\int_{x_1}^{x_2}{\d x\over\sqrt{\lambda^2-u^2}}.
\]
 This integral suggests the substitution $u=\lambda\sin\theta$ so
\[
T_r={1\over\lambda}\int_{x_1}^{x_2}{\d x\over\cos\theta}.
\]
 \Henon\ now introduces 
 \[\label{eq:defsF}
F(u)\equiv{\d x\over\d u}
\]
 so $\d x=F(u)\,\d u= F(u)\lambda\cos\theta\,\d\theta$ and $T_r$ becomes
 \[\label{eq:Tr2}
T_r=\int_{-\pi/2}^{\pi/2}\d\theta\, F(\lambda\sin\theta).
\]
 In terms of the Maclaurin expansion of $F(u)$
 \[
F(u)=a_0+a_1u+a_2u^2+\cdots,
\]
 equation (\ref{eq:Tr2}) becomes
 \[
T_r=a_0\pi+a_2\lambda^2{\pi\over2}+a_4\lambda^4{3\pi\over8}+\hbox{O}(\lambda^6);
\]
 none of  the odd-numbered terms in the series makes a contribution to $T_r$.
If $T_r$ is to be independent of $L$, it must be independent of $\lambda$
(eq.~\ref{eq:defslam}), so the function $F$ must have vanishing coefficients
$a_2,a_4,\ldots$: its even part must consist only of the constant $a_0$,
which specifies $T_r$ through
 \[
T_r=a_0\pi.
\]

Integrating equation (\ref{eq:defsF}) we obtain
 \[
x=\int\d u\,F(u)=a_0u+G(u^2)={T_ru\over\pi}+G(u^2),
\]
 where $G(u^2)$ is the even function of $u$ that is obtained by integrating
the odd part of $F$. If we subtract the values taken by this expression  at
$x=x_2$ and $x=x_1$, we obtain
 \[\label{eq:d2}
x_2-x_1={2T_r\lambda\over\pi}
\]
 because at these points $u=\pm\lambda$. Squaring this expression and
 recalling the definition (\ref{eq:defslam}) of $\lambda$, we have
\[\label{eq:d3}
(x_2-x_1)^2={4T_r^2\lambda^2\over\pi^2}=t(\Lc^2-L^2).
\]
 where
\[
t\equiv {4T_r^2\over\pi^2}.
\]

The points $x_1$, $x_2$ are functions of both $E$ (which determines the
slopes of the lines in \figref{fig:one} and $L^2$ (which determines the $y$
intercepts of these lines). We next differentiate our expressions
(\ref{eq:d1}) and (\ref{eq:d3}) with respect to first $E$ and then $L^2$:
 \begin{eqnarray}\label{eq:set}
x_i+E{\pa x_i\over\pa E}-{\d f\over\d x}{\pa x_i\over\pa E}&=&0\nonumber\\
2(x_2-x_1)\left({\pa x_2\over\pa E}-{\pa x_1\over\pa E}\right)&=&
(\Lc^2-L^2){\d t\over\d E}+t{\d\Lc^2\over\d E}\nonumber\\
E{\pa x_i\over\pa L^2}-1-{\d f\over\d x}{\pa x_i\over\pa L^2}&=&0\\
2(x_2-x_1)\left({\pa x_2\over\pa L^2}-{\pa x_1\over\pa
L^2}\right)&=&-t.\nonumber
 \end{eqnarray}
 We use the first of these equations to eliminate $\pa x_i/\pa E$ from the
 second equation. This operation yields
 \[\label{eq:nearly}
2(x_2-x_1)\left({x_2\over f_2'-E}-{x_1\over f_1'-E}\right)=
(\Lc^2-L^2){\d t\over\d E}+t{\d\Lc^2\over\d E}\\
\]
 where $f_2'\equiv\d f/\d x|_{x_2}$, etc. But from the third of equations
 (\ref{eq:set}) we have 
\[
{\pa x_i\over\pa L^2}={1\over E-f_i'},
\]
 so equation (\ref{eq:nearly}) can be written
\[
-2(x_2-x_1)\left(x_2{\pa x_2\over\pa L^2}-x_1{\pa x_1\over\pa L^2}\right)=
(\Lc^2-L^2){\d t\over\d E}+t{\d\Lc^2\over\d E}.
\]
 Between this  equation and the last of equations  (\ref{eq:set}) we can sove
 for the individual derivatives of $x_1$ and $x_2$:
\begin{eqnarray}
2(x_2-x_1)^2{\pa x_2\over\pa L^2}&=&(\Lc^2-L^2){\d t\over\d
E}+t{\d\Lc^2\over\d E}+\fracj12x_1t\nonumber\\
2(x_2-x_1)^2{\pa x_1\over\pa L^2}&=&(\Lc^2-L^2){\d t\over\d
E}+t{\d\Lc^2\over\d E}+\fracj12x_2t.
\end{eqnarray}
 Adding these equations, we obtain
\[
(x_2-x_1)^2{\pa(x_1+x_2)\over\pa L^2}=
(\Lc^2-L^2){\d t\over\d E}+t{\d\Lc^2\over\d
E}+\fracj12t(x_1+x_2).
\]
 We use equation (\ref{eq:d3}) to eliminate $(x_2-x_1)^2$ from this equation
 and rearrange the result:
\[
{\pa(x_1+x_2)\over\pa L^2}-{x_1+x_2\over2(\Lc^2-L^2)}=
-{1\over t}{\d t\over\d E}-{1\over\Lc^2-L^2}{\d\Lc^2\over\d E}
\]
 This is a first-order, linear differential equation. Its integrating factor
is $\sqrt{\Lc^2-L^2}$, so its general solution follows from
\begin{eqnarray}
(x_1+x_2)\sqrt{\Lc^2-L^2}&=&-{\d\ln t\over\d E}\int\d\Lc^2\,
(\Lc^2-L^2)^{1/2}-{\d\Lc^2\over\d E}\int{\d
L^2\over\sqrt{\Lc^2-L^2}}\nonumber\\
&=&{\d\ln t\over\d E}\fracj23
(\Lc^2-L^2)^{3/2}+{\d\Lc^2\over\d E}2\sqrt{\Lc^2-L^2}+\hbox{const}
\end{eqnarray}
 Since a non-zero value of the constant of integration would cause $x_1+x_2$
 to diverge as $L\to\Lc$, it must be set to zero, and we have finally
\[\label{eq:xpx}
x_1+x_2=\fracj23{\d\ln t\over\d E}
(\Lc^2-L^2)+2{\d\Lc^2\over\d E}.
\]
 This equation specifies as a function of the angular momentum $L$ the
distance in \figref{fig:one} between the points of intersection of the
potential's curve and the straight line. Hence it implicitly specifies the
curve and thus the potential. To tease out that specification we first write down
the quadratic equation that has roots at $x_i$:
\[
[2x-(x_1+x_2)]^2=(x_2-x_1)^2
\]
 and use equations (\ref{eq:xpx}) and (\ref{eq:d3}) to eliminate $x_1+x_2$
 and $x_2-x_1$:
\[
\left[2x-\fracj23{\d\ln t\over\d E}
(\Lc^2-L^2)-2{\d\Lc^2\over\d E}\right]^2=t(\Lc^2-L^2).
\]
 Now we eliminate $L^2=Ex_i-f(x_i)$ (eq.~\ref{eq:d1})
\[
\left[2x-\fracj23{\d\ln t\over\d E}
(\Lc^2-Ex_i+f(x_i))-2{\d\Lc^2\over\d E}\right]^2=t(\Lc^2-Ex_i+f(x_i)).
\]
 This equation holds only when $x=x_i$, so we set it to that value and
 rearrange to
\begin{eqnarray}\label{eq:parab1}
\Bigl[\fracj23{\d\ln t\over\d E}f_i
&-&\left(2+\fracj23E{\d\ln t\over\d E}\right)x_i
+\left(\fracj23\Lc^2{\d\ln t\over\d E}+2{\d\Lc^2\over\d
E}\right)\Bigr]^2\nonumber\\
&=&t(f_i+\Lc^2-Ex_i).
\end{eqnarray}
 Since $(x_i,f_i)$ is the locus of a point on the potential's curve, we now
have the explicit equation of that curve. The coefficients $\d\ln t/\d E$ etc
that form this relation between $x_i$ and $f_i$ are functions of $E$, but the
curve specified by the relation does not depend on $E$; $E$ specifies the
slope of the straight lines along which individual points $(x_i,f_i)$ lie,
not the curve itself.

The standard equation of a parabola
is 
\[
y^2=4ax.
\]
 When we make a general linear transformation,
 \[
\pmatrix{x\cr y}=\pmatrix{\alpha&\beta\cr \gamma&\delta}
\pmatrix{x'\cr y'}+\pmatrix{x_0\cr y_0}
\]
 this becomes
\[\label{eq:parab2}
(\delta y'+\gamma x'+y_0)^2=4a(\alpha x'+\beta y'+x_0)
\]
 This equation has the same form as our equation (\ref{eq:parab1}) for the
curve of the potential, so requiring $T_r$ to depend only on $E$ implies
that the curve of $r^2\Phi$ versus $r^2$ is a parabola.
$\Phi$ may diverge at the centre, but it will not do so more strongly than
$1/r$, so $f=2r^2\Phi\to0$ as $r\to0$, and the origin always lies on the
potential's parabola. Moreover
 \[
{\d f\over\d x}={\d(x\Phi)\over\d x}=\Phi+x{\d\Phi\over\d x}
\]
 so the gradient of the parabola at the origin is the central value of $\Phi$.
Also, the gradient of the chord from the origin to an arbitrary point $(x,f)$ on
the parabola is $f/x=\Phi(r)$; in particular, the value of the potential at
infinity is given by the parabola's asymptotic gradient.

To avoid the complex expressions in (\ref{eq:parab1}), we reformulate the
expression in the form (\ref{eq:parab2})
\[\label{eq:parab3}
(\delta f+\gamma x+y_0)^2=4a(\alpha x+\beta f)+y_0^2,
\]
 where the constant on the right has to been chosen to ensure that the origin
lies on the parabola. As $x\to\infty$, $f=x\Phi$ will grow too, and the left
side of the equation will grow faster than the right side unless $\delta f$ tends
to $\gamma x$. Hence
 \[
\delta\Phi_\infty=-\gamma
\]
 and we may rewrite (\ref{eq:parab3}) in the form
\[\label{eq:parab4}
[\delta(\Phi-\Phi_\infty)x+y_0]^2=4ax(\alpha+\beta \Phi)+y_0^2
\]
 If we choose to set the zero point of $\Phi$ at $x=\infty$,
this  quadratic equation for $\Phi(x)$ becomes
 \[
\delta^2x\Phi^2+(2\delta y_0-4a\beta)\Phi-4a\alpha=0,
\]
 so
 \[
\Phi(x)={(4a\beta-2\delta y_0)\pm\sqrt{(2\delta
y_0-4a\beta)^2+16\delta^2x
a\alpha}\over2\delta^2x}
\]
 We cast this into a more familiar form by multiplying top and bottom by the
 top with the optional sign reversed
 \[
\Phi(x)=-{8a\alpha
\over(4a\beta-2\delta y_0)\mp\sqrt{(2\delta
y_0-4a\beta)^2+16\delta^2xa\alpha}}
\]
 The plus sign must be chosen to ensure that  the central potential is
finite. Then setting 
\[
GM\equiv{\sqrt{2a\alpha}\over\delta}\quad
\hbox{and}\quad
b={a\beta/\delta-\fracj12y_0\over\sqrt{2a\alpha}}
\]
we obtain the
isochrone potential in its classic form
 \[
\Phi(r)=-{GM\over b+\sqrt{b^2+r^2}}.
\]

\Henon\ plots the potential and its radial force out to $5b$. He gives the
formula for the mass contained within radius $r$ and, taking the derivative
of this formula, gives the generating density $\rho(r)$. He remarks that the
density is everywhere positive and that $\rho\sim r^{-4}$ at large $r$. He
plots the projected density out to $r=5b$ together with data from five
globular clusters, and the model is seen to fit the data to within the
variation between clusters.

Then he asks ``why do clusters resemble the isochrone?''. He remarks that
perturbations in the cluster potential due to individual stars will cause the
cluster's density profile to evolve until some stable configuration is
reached. Normally the endpoint of this evolution is taken to be when the
velocity distribution has become Maxwellian, but this distribution is
only possible in the isothermal sphere, which necessarily has infinite mass.

\Henon\ argues that stars that share the same radial period are resonantly
coupled, so they will share energy on a shorter timescale than stars that
have different radial periods. In any other model than an isochrone, the set
of resonantly coupled stars will contain stars that differ in energy.
Consequently these stars will exchange energy. \Henon\ hypothesises that
in these exchanges stars with less energy will gain energy from those with
more energy. He further hypothesises that these energy changes will cause the
mass distribution to evolve towards the isochrone model. Once this distribution has
been reached, net energy exchanges between resonantly coupled stars will
cease, and thus the cluster will cease to evolve.

\Henon\ recognises that a satisfactory exploration of this idea is very hard
and proposes a preliminary test of concept. To develop such a test he
considers ``hyperbolic models'' in which the curve $f(x)$ is a hyperbola
rather than a parabola. For these models at fixed $E$, $T_r$ decreases with
increasing $L$, with the consequence that at given $E$, circular orbits have
shorter periods than eccentric orbits. Conversely, in the set of resonantly
coupled stars, circular orbits have more energy than eccentric orbits. So in
a hyperbolic model eccentric orbits will gain energy from from circular orbits. It
follows that the circular orbits will shrink while eccentric orbits grow in size.
But in any cluster the outer regions are dominated by eccentric orbits, so
resonant interactions in a hyperbolic model will enhance the density in the
outer regions, reducing the central concentration of the model. A plot of the
density profiles of the isochrone and a hyperbolic model shows that the
latter is more centrally concentrated than the former. Therefore resonant
interactions in the latter will drive the hyperbolic model in the direction
of the isochrone.

In the concluding section \Henon\ remarks that whatever the validity of the
hypothesis of resonant interactions, the isochrone constitutes a realistic
cluster model for which all quantities are analytically available.

\section{Isochrone papers two and three}

The second paper \citep{HenonII} computes orbits in the isochrone.  \Henon\ uses
$U=1-\Phi/\Phi(0)$ as the radial coordinate, explicitly integrating $\d U/\d
t=(\d U/\d r)\dot r$ so obtain both $t(U)$ and the angular coordinate
$\psi(U)$. He simplifies these formulae by introducing the potentials $U_1$
and $U_2$ at peri- and apo-centre. He tabulates $U_i$, $r_i$, $T_r$ and the
mean density along the orbit for several values of $E$ and $L$. He plots a
few of these orbits.

The third paper \citep{HenonIII} uses Eddington's inversion formula to
compute the isochrone's distribution function (\df) $f(E)$. \Henon\ tabulates
both $f(E)$ and $\d N/\d E$, the number of stars with a given energy.

\section{Why is the isochrone important?}

From the current perspective, the isochrone is one member of a family of
useful spherical systems: the other members are the \cite{Jaffe} model and the
\cite{Hernquist} model. These models are all derived from
simple functional forms of $\Phi(r)$:
\begin{eqnarray}
\Phi_{\rm Jaffe}(r)&=&-4\pi G\rho_0b^2\ln(1+b/r)\nonumber\\
\Phi_{\rm Hern}(r)&=&-2\pi G\rho_0b^2/(1+r/b),
\end{eqnarray}
 and for each one has analytic forms of both the density $\rho(r)$ and
the ergodic \df\ $f(E)$. What uniquely distinguishes the
isochrone is the availability of analytic formulae for the actions and
angles of its orbits. This availability is implicit in \Henon's second paper
but he does not mention this. The relevant formulae appear to have been first
given by \cite{Saha}.

\subsection{Role of the isochrone in torus mapping}

Regular orbits are those whose time series $x(t)$, $z(t)$, etc., are
quasiperiodic: when these time series are Fourier transformed, all the
frequencies that occur can be expressed as integer linear combinations of
three fundamental frequencies. \cite{Arnold} shows that a quasiperiodic orbit
admits three isolating integrals, which confine the orbit to a three-torus in
six-dimensional phase space. These tori are null in the sense that the
Poincar\'e invariant $\sum_i\int\d x_i\d v_i$ of every two surface in the
torus vanishes. The natural labels of the tori are the actions
\[
J_i={1\over2\pi}\oint_{\gamma_i}\vv\cdot\d\vx
\]
 where for $i\ne j$ $\gamma_i$ and $\gamma_j$ are two closed paths around the
torus that cannot be distorted into one another while staying within the
torus. These are the only labels of individual tori that can be complemented
by canonically conjugate variables $\theta_i$, the ``angle'' variables. The
$\theta_i$ specify positions with a torus.

If a canonical transformation is used to map a null torus, it remains a null
torus, and if the image torus can be restricted to a constant-energy surface,
it becomes an orbital torus: a surface to which an orbiting particle is
confined by motion under the Hamiltonian.  The idea behind torus mapping is
to arrange for the Hamiltonian to be constant on a null torus of
pre-determined actions by adjusting the parameters of a canonical
transformation such that $H(\vx,\vv)=\hbox{constant}$ on the image torus
\citep{McGillB}. If phase space -- or a non-negligible part of it -- can be
foliated with orbital tori in this way, the images of the angle-action
coordinates of the analytic tori become angle-action coordinates of the
Hamiltonian $H$.  Any analytic tori can be used for this purpose, but in
practice one tries to use tori that are as similar as possible to the orbital
tori of $H$. In this regard the isochrone potential is well suited to mapping
tori into orbital tori for many galactic potentials, and it was the choice of
\cite{McGillB}.  The harmonic-oscillator potential (which is really a
limiting case of the isochrone potential) has also been used for torus
mapping \citep{KaasalainenB94a}. 

If the Hamiltonian $H$ is not integrable (i.e., many of its orbits are not
quasiperiodic) it will prove impossible to arrange for $H$ to be constant on
the image torus for some or all action values. Nevertheless, one can foliate
phase space with the image tori that result from minimising the variance of
$H$ over the image torus. Then if we define $\overline{H}(\vJ)$ to be the
angle-averaged value of $H$ on the image torus with actions $\vJ$,
$\overline{H}$ becomes an integrable Hamiltonian with explicitly known
angle-action coordinates, and the difference
$\Delta(\vx,\vv)=H(\vx,\vv)-\overline{H}$ becomes a small perturbation of
this integrable Hamiltonian that yields the original Hamiltonian. Thus torus
mapping allows us to study motion in a general Hamiltonian as perturbation of
a very close integrable Hamiltonian \citep{KaasalainenB94b}.

For over a decade after its introduction, torus mapping found little
application. In the last several years it has proved a valuable tool for
modelling our Galaxy \citep{McMillanB08,McMillan,McMillanB13} and I believe
it will play a significant role in the scientific exploitation of the
billion-Euro Gaia survey.

\begin{figure}
\includegraphics[width=.99\hsize]{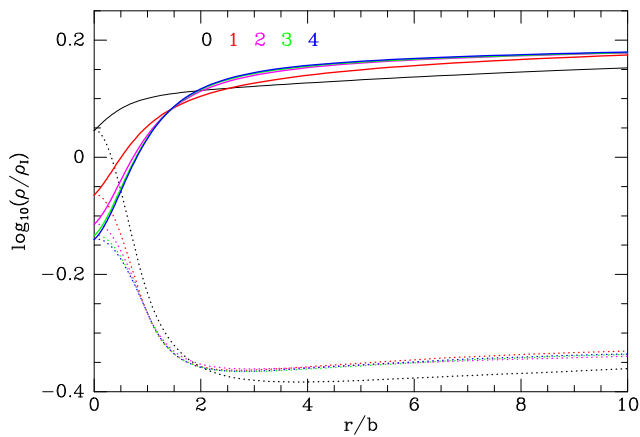}
\caption{Convergence of a flattened isochrone model model.
We plot  $\log_{10}[\rho(r,\theta)/\rho_{\rm
I}(r)]$, where $\rho_{\rm I}(r)$  is the density of the isochrone sphere, for
two values of the colatitude $\theta$: full curves are for a ray that lies close to the
major axis while dotted curves are for a ray that lies close to the minor
axis. The colour of the curves indicates which iterate of the potential was
used for the density evaluation. The \df\ of this model is obtained from that
of the isochrone sphere by replacing $J_z$ with $1.4J_z$ and $J_\phi$ by
$0.7J_\phi$. (From  Binney 2014)}
\label{fig:0itr}
\end{figure}

\subsection{Flattened isochrones}

Since \Henon\ gave is the isochrone's \df\ $f(H)$ and we know how to write
$H$ as a function of the actions $\vJ$, it is trivial to express the
isochrone's $f$ as a
function of $\vJ$.

Since angle-action coordinates are canonical, the Jacobian between these
coordinates and ordinary $(\vx,\vv)$ coordinates is unity, and the density of
stars in angle-action space is $f(\vJ)$. Moreover, the range of each angle
coordinate is always $(0,2\pi)$, so the phase-space volume occupied by orbits
with actions in $\d^3\vJ$ is $(2\pi)^3\d^3\vJ$ and thus the density of stars
in three-dimensional action space is $(2\pi)^3f(\vJ)$. That is, to within an
uninteresting constant factor, the function $f(\vJ)$ that completely
characterises the dynamics of a stellar population is the density of stars in
a readily imagined three-dimensional space.

In an ergodic model such as \Henon's isochrone, the action-space density of
stars is constant on surfaces $H(\vJ)=\hbox{constant}$, which are roughly
triangular surfaces $\vk\cdot\vJ=\hbox{constant}$, where $\vk$ is a vector
whose direction is determined by $H$. In this ergodic model all three
velocity dispersions are equal.

If we shift stars over surfaces of constant $H$, so the action-space density
of stars becomes non-uniform on these surfaces, we generate a model that has
velocity anisotropy: if we shift the stars towards the $J_r$ axis, we
generate radial anisotropy. If we shift the stars away from the $J_z$ axis,
we flatten the model. So long as we leave unchanged the number of stars on each
constant-$E$ surface, the radial density profile is essentially unchanged:
the model may become elongated, or flattened, and it may develop velocity
aniotropy, but it retains essentially the same radial density profile.

A significant advantage of considering the \df\ to be a function of the
actions rather than a function of energy and other isolating integrals is
that it is then very easy to find the gravitational potential of the
self-consistent model that has the given \df. One simply guesses a potential
$\Phi_0$ and on a spatial grid evaluates the density
$\rho_{1/2}(\vx)=\int\d^3\vv\,f(\vJ)$ implied by the \df\ in that potential.
One then solves Poisson's equation for the corresponding potential
$\Phi_{1/2}(\vx)$ at the grid points. Then one takes as a new guess of the
self-consistent potential
\[
\Phi_1(\vx)=(1+\gamma)\Phi_{1/2}(\vx)-\gamma\Phi_0(\vx)
\]
 with $\gamma\simeq0.5$, and repeats the process, which converges after 3--5
iterations -- see Figure \ref{fig:0itr}.

A prerequisite for this program is the ability to evaluate the action
integrals given an arbitrary point $(\vx,\vv)$. A very convenient technique,
which is remarkably accurate for modified isochrone models, is the ``St\"ackel
Fudge'' of \cite{Binney12a}, and this is the technique used by
\cite{Binney14} to explore flattened isochrone models.

There are several ways in which one can change an ergodic \df\ into the \df\
of a flattened model and it remains unclear what the best approach is.
\cite{Binney14} replaced each action $J_i$ in the ergodic \df\ by
$\alpha_iJ_i$. If, for example, $\alpha_r>1$, the new \df\ decreases with
increasing $J_r$ more rapidly than in the ergodic \df\ and the model becomes
tangentially biased, and conversely if $\alpha_r<1$. If $\alpha_z>1$, motion
perpendicular to the equatorial plane is being discouraged, and the model
becomes flattened.

\section{Conclusions}

\Henon's isochrone is a fine example of the kind of curiosity-driven research
that is so much discouraged by the current funding environment. There is no
way that \Henon\ could have justified to a grants committee his effort to
find the most general potential that made $T_r$ independent of angular
momentum. I suppose he was first just curious, and then became fascinated by
the puzzle posed by the search for the isochrone. As Section 1 shows, solving this
puzzle required quite a tour de force, and was surely not done in some spare
afternoon.

Once \Henon\ had found the isochrone, he was impressed by how well it
represented globular clusters. And rightly so, because it is really very
surprising that a potential chosen to have orbits with a particular property
should even have a non-negative generating density; that it also have a
non-negative \df\ is improbable; that it should also provide an excellent fit
to globular clusters is astounding. 

\Henon's attempt at a physical explanation of the closeness of globular
clusters to the isochrone is highly ingenious and shows deep physical
insight.  Within the last decade resonant relaxation has become fashionable
in studies of galactic nuclei and of planetary systems, but in 1958 \Henon's
use of the idea was ground-breaking. That said, I don't think it was
convincing. Yes stars with common values of $T_r$ can more readily exchange
energy than stars with unrelated frequencies, but it is far from clear that
the sense of that exchange will be to establish equipartition of energy: the
essence of the gravithermal catastrophe is that in self-gravitating systems,
stars with less energy lose energy to those with more, so in self-gravitating
systems wealth inequalities grow as fast as they do just now in the United
States. Moreover, why the focus on stars that have equal $T_r$ rather than
$T_\phi$?

\Henon\ was recognised that he was only scratching the surface of how a
globular cluster might evolve through resonant relaxation. I think this
problem remains an intriguing issue in dynamics, and one that might now be
addressed with the help of techniques, torus mapping and taking actions
as the arguments of \df s, that make extensive use of \Henon's delightful
isochrone.

\section*{acknowledgement} 

This work has been supported by STFC by grants R22138/GA001 and ST/K00106X/1
and by the European Research Council under the European Union's Seventh
Framework Programme (FP7/2007-2013) / ERC grant agreement no.\ 321067.

\end{document}